\documentclass[aip, apl, graphicx, groupedaddress, numerical, reprint]{revtex4-1}

\usepackage{graphicx}

\draft 

\begin{document}


\title{Photo-induced doping and strain in exfoliated graphene} 



\author{E. Alexeev}
\email{E.Alexeev@exeter.ac.uk}
\author{J. Moger}
\author{E. Hendry}
\affiliation{School of Physics, University of Exeter, Stocker Road, Exeter, EX4 4QL, United Kingdom}


\date{\today}

\begin{abstract}
The modification of single layer graphene due to intense, picoseconds near-infrared laser pulses is investigated. We monitor the stable changes introduced to graphene upon photoexcitation using Raman spectroscopy. We find that photoexcitation leads to both a local increase in hole doping and a reduction in compressive strain. Possible explanations for these effects, due to photo-induced oxygenation and photo-induced buckling of the graphene, are discussed. 
\end{abstract}


\maketitle 

The unique electrical and optical properties of graphene have made it a very promising material for future electro-optical applications. One of graphene most appealing aspects, due to its interfacial nature, is its tunability. For example, it is well known that the band structure and conduction properties of graphene can be modified and functionalized by molecular adsorbates\cite{01Wehling2008, 02Wehling2008, 03Wang2010, 04Chen2007, 05Robinson2010}, by irradiation under electron beams\cite{06Teweldebrhan2009, 07Fischbein2008, 08Withers2011, 09Sessi2009}, by applied electric and magnetic fields\cite{10Castro2007, 11Kane2005, 12McCann2006, 13Zhang2009} and by nano-structuring of the material\cite{14Han2007, 15Tapaszto2008, 16Stampfer2009, 17Lemme2009, 18Hong2012}. Indeed, the ability to control the majority carrier type while introducing a band gap makes graphene promising for nanocircuit design\cite{19Li2008,20Wang2008,21Berger2004,22Lin2011}.

Graphene has also been shown to demonstrate interesting behaviour under optical illumination. For example, novel photochemical approaches have been developed to achieve efficient graphene modification and bandgap modulation.  In Ref.~\onlinecite{23Luo2012} ultraviolet radiation was shown to induce doping of the irradiated areas of CVD-grown graphene, with no significant reduction of the carrier mobility. Thus, photo-modification is an efficient means by which to create channels with increased conductivity, forming in-built electrodes. Moreover, by covering the graphene with a layer of fluoropolymer, it has been shown\cite{24Lee2012} that irradiated areas become fluorinated, which, in turn, leads to a significant increase of resistivity in these regions. On increasing photoexcitation intensity, graphene also exhibits ablation\cite{25Stoehr2011}, which can be used to create complex structures within single graphene flakes. This approach is applicable on suspended samples and therefore is advantageous over standard etching techniques. The authors of Ref.~\onlinecite{25Stoehr2011}, for example, managed to fabricate graphene dots with diameters less than 100~nm and nanoribbons down to 20~nm in width. 

In this paper we report the photomodification effects of near-infrared, picosecond laser radiation on exfoliated graphene flakes. Using Raman spectroscopy as a probe, we observe that, for laser irradiation well below the damage threshold of graphene, the photo-interaction leads to changes in both doping and strain in the graphene flake. The localisation and stability of the introduced changes make it suitable for future patterning applications.

Single layer graphene samples were prepared by mechanical exfoliation of natural graphite and deposited on 100-$\mu$m-thick glass substrates. The number of layers in individual samples were estimated by optical contrast measurements\cite{26Gaskell2009} and confirmed using Raman spectroscopy\cite{27Ferrari2006}, allowing isolation of monolayer flakes. 

Photomodication and photoluminescence microscopy were performed using a custom-build non-linear optical microscope based on a commercial inverted microscope and confocal laser-scanning unit (IX71 and FV300, Olympus~UK). A full description of the system can be found in Ref.~\onlinecite{27Moger2008}. Picosecond excitation was provided by an optical parametric oscillator (OPO), (Levante Emerald, APE Berlin) pumped with a frequency doubled Nd:Vandium picosecond oscillator (High-Q Laser Production GmbH). The signal beam from the OPO was used to generate excitation pulses centred at 816~nm with a width of 6~ps and repetition rate of 76~MHz. 

\begin{figure}
\includegraphics{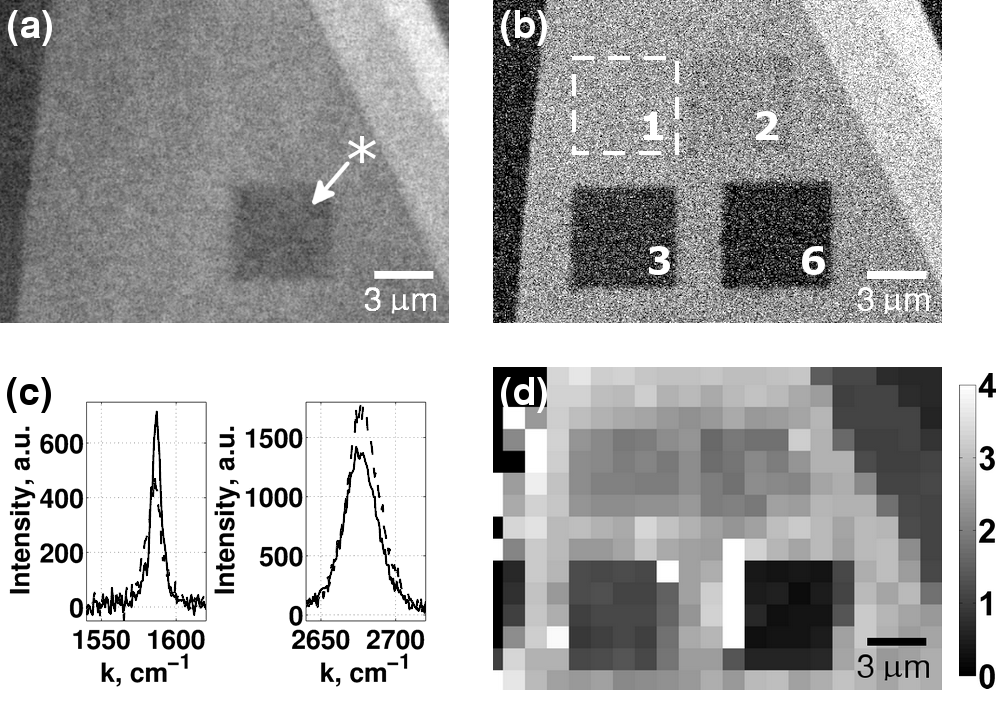}
\caption{\label{fig1}(a) Optical micrograph of monolayer graphene flake after phomodification. Asterisk denotes the area where pump fluence was high enough to cause material ablation. (b) Photoluminescence image of the same flake showing the decrease of up-converted photoluminescence intensity in modified regions. Numbers indicate excitation fluence in  $\textrm{mJ}/\textrm{cm}^2$ that was used to modify selected region. Dashed line indicates the first excited region. (c) Raman spectra corresponding to the centre of the first square region (dashed) before and (solid) after photoexcitation with 1  $\textrm{mJ}/\textrm{cm}^2$ laser pulses for 1 minute (d) Raman map of the sample plotting intensity ratio of the 2D and G peaks. Decrease of intensity ratio in modified regions indicates an increased level of doping.}%
\end{figure}

Photoluminescence imaging and optical modification were performed using a 60X, 1.2~NA water immersion objective (UPlanS Apo, Olympus UK) to focus a diffraction limited spot onto the sample which could be raster-scanned over designated areas of the sample. Up-converted photoluminescence, isolated with a 750~nm short-pass filter (FES0750, Thorlabs), was monitored using a photomultiplier tube and used for sample imaging, as discusses in Ref.~\onlinecite{28Stoehr2010} - see Fig.~\ref{fig1}~(b). The exposure time of the sample excitation was controlled by varying the number of raster-scans.

In order to minimize changes induced to the sample during imaging, laser fluence was kept below 0.2~$\textrm{mJ}/\textrm{cm}^2$. For the photomodification, a laser beam with the fluence in the range of 1~-~3~$\textrm{mJ}/\textrm{cm}^2$ was raster scanned over the chosen area of the monolayer flake. Raman spectra were subsequently collected in a separate, commercial Raman spectrometer (RM1000, Renishaw), using excitation beam with a wavelength of 532~nm and intensity of 5~mW that was focused to a spot size of 1.5~$\mu$m. All measurements were performed at room temperature and in ambient air. 

To investigate effects of photoexcitation square regions of monolayer flakes were exposed to different fluences of picosecond excitation for a duration of 1 minute. Figure~\ref{fig1}~(a) shows optical micrograph of a single layer graphene flake deposited on a glass substrate after photoexcitation. Even though the contrast of the image has been artificially increased to make the monolayer part visible, there is no visible sign of modification until the onset of ablation (bottom right square in Fig.~\ref{fig1}~(a), (b) and (d)). However, the square areas that have been exposed to the laser excitation can be clearly seen in the photoluminescence images, Fig.~\ref{fig1}~(b), as a clear reduction in photoluminescence intensity. The reduction in photoluminescence increases with excitation fluence , and the induced changes were found to be stable over the duration of the project (i.e. several months). However, we found that they could be reversed by immersing sample in a solvent, such as methanol or isopropyl alcohol, for one hour. A possible explanation for the photoluminescence intensity decrease could lie in photo-induced changes of carrier concentration or relaxation time. Since photoluminescence originates from non-equilibrium distribution of photoexcited carriers, it is very sensitive to the carrier relaxation dynamics. Recent studies show that charge doping of graphene flakes leads to changes of photoexcited carriers relaxation due to the carrier heating effect\cite{37Tielrooij2013, 38Jnawali2013}. However, the precise mechanism for this effect on photoluminescence is not fully understood, and is to be the focus of future investigation. Here we focus on the changes induced to the graphene itself, modified by exposure to the picoseconds laser pulses. It is worth noting that it was not possible to observe similar modification effects with femtosecond excitation, due to low damage threshold for such ultrafast pulses\cite{29Currie2011}. The duration of the picoseconds pulses, meanwhile, is comparable in length to the lattice cooling timescales of graphene flakes\cite{30Hale2011}, allowing for efficient heating without damaging the graphene flakes.

In order to understand the changes introduced in graphene by laser irradiation we use Raman spectroscopy. Figure~\ref{fig1}~(c) shows Raman spectra of the point corresponding to the centre of the first modified region before (dashed) and after (solid) photomodification by 1~$\textrm{mJ}/\textrm{cm}^2$ laser pulses for 1~minute. The G peak at 1580~$\textrm{cm}^{-1}$ originates from the doubly degenerate $E_{2g}$ phonon mode at the Brillouin zone centre, while the 2D peak at 2700~$\textrm{cm}^{-1}$ corresponds to a double-resonance process, involving two transverse optical phonons near the K point. A symmetric 2D peak with a width of 25-45~$\textrm{cm}^{-1}$ is characteristic of monolayer graphene. Upon photoexcitation, the G peak is up-shifted by 1.2~$\textrm{cm}^{-1}$ and the 2D peak is down-shifted by 1.6~$\textrm{cm}^{-1}$, and the intensity ratio of two peaks $I_{2D}/I_G$ is decreased. Note that we have not observed the defect-induced D peak at 1350~$\textrm{cm}^{-1}$, which indicates that photoexcitation does not induce structural defects.  In Fig.~\ref{fig1}~(d), we plot a Raman map of the intensity ratio of the 2D to G peaks for the flake. The photomodified square areas can be clearly seen, which indicates that modification is local, limited to the region of photoexcitation. Higher laser irradiation intensity gives rise to the stronger decrease of the intensity ratio, which corresponds to higher level of doping\cite{31Das2008}.

\begin{figure}
\includegraphics{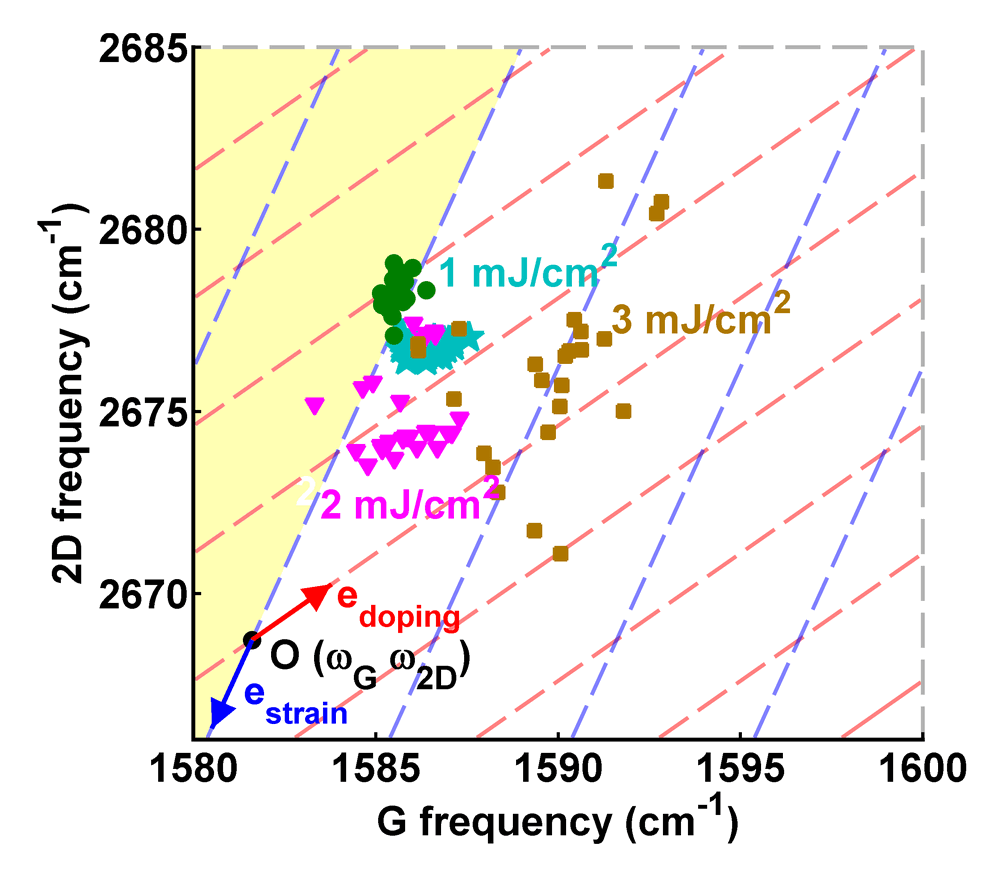}
\caption{\label{fig2}(Color online) Plot of correlated peak position after photomodification with 1~$\textrm{mJ}/\textrm{cm}^2$ (cyan stars), 2~$\textrm{mJ}/\textrm{cm}^2$ (magenta triangles) and 3~$\textrm{mJ}/\textrm{cm}^2$ (brown squares) for one minute. Position for the sample before photoexcitation is shown by green circles. Black dot in the bottom left corner denotes $(\omega_G ; \omega_{2D})$ position not affected by strain or charge doping. Arrows indicate directions of strain- and doping-induced movement of $(\omega_G ; \omega_{2D})$.}%
\end{figure}

It is well known that changes of the Raman 2D and G peak positions and intensities can be caused by both changes in doping and strain, and this bimodal sensitivity complicates data analysis\cite{31Das2008, 32Mohiuddin2009, 33Ni2008, 34Zabel2012}. However, this can be overcome by considering their correlated position. To get a qualitative description of the changes induced in graphene upon photoexcitation we apply the analysis first introduced in  Ref.~\onlinecite{35Lee2012}. This analysis is based on the fact that the fractional variation of peak positions  $\Delta \omega_{2D}/\Delta \omega_{G}$ is very different for cases of strain and doping. The average value of $\Delta \omega_{2D}/\Delta \omega_{G}$ for uniaxial strain of random direction is 2.2, while for the carrier concentration above $1.4 \times 10^{12} \textrm{cm}^{-2}$ for the case of hole doping (i.e. a down-shift of the Fermi level) $\Delta \omega_{2D}/\Delta \omega_{G}$ is approximately 0.7. Therefore contributions of strain and doping to the correlated peak position $(\omega_G ; \omega_{2D})$ can be separated using vector decomposition with the gradient for unit vectors for strain- and doping-induced changes being 2.2 and 0.7, respectively. It should be noted that such a vector decomposition is more complicated for electron doping of graphene; however, our experiments on contacted graphene flakes revealed an increased level of hole (p) -doping as a result of photoexcitation. The origin for the correlated position plots, i.e.  $(\omega_G ; \omega_{2D})$ in Fig.~\ref{fig2} can be obtained from the results of Ref.~\onlinecite{35Lee2012} using Raman peak dispersion\cite{27Ferrari2006} to take into account different Raman excitation wavelength, giving $(\omega_G ; \omega_{2D})= (1581.6 \textrm{cm}^{-1}; 2668.7 \textrm{cm}^{-1})$.

Figure~\ref{fig2} shows representative example of correlated peak position for different areas of the sample after photomodification using different laser fluences. Dashed blue (red) lines indicate direction of strain (doping) induced movement of the $(\omega_G ; \omega_{2D})$ point for different constant values of strain (doping). The $(\omega_G ; \omega_{2D})$ upshift (downshift) from the origin along 'strain' lines corresponds to increasing compressive (tensile) strain. The $(\omega_G ; \omega_{2D})$ upshift (downshift) along 'doping' lines corresponds to increasing (decreasing) p-doping. The figure also shows origin (black dot) and unit vectors for doping (red) and strain (blue) induced peak shifts that can be used for vector decomposition. The yellow shaded region indicates a 'forbidden area': since increasing doping leads to up-shift of the correlated peak position from the origin, $(\omega_G ; \omega_{2D})$ can enter this area only for low levels of doping ($< 1.4 \times 10^{12} \textrm{cm}^{-2}$) when its dependence on Fermi level position becomes nonlinear.
Since the native strain leads to non-negligible variation of  $(\omega_G ; \omega_{2D})$, we need to take into account peak positions for the pristine sample. The green circle markers denote correlated peak position of non-modified sample. They form a narrow group with primarily strain-induced variation. However, the distribution changes dramatically after photomodification. After excitation with 1~$\textrm{mJ}/\textrm{cm}^2$ (cyan stars) laser light the centre of distribution is shifted down and to the right hand side. To achieve this kind of movement, $(\omega_G ; \omega_{2D})$ should be up-shifted along 'doping' line and down-shifted along 'strain' line. The former indicates the increase of the local doping level. The similar effect was observed in Ref.~\onlinecite{23Luo2012} for UV excitation, where changes of the doping level were attributed to the photo-induced release of electron trapping adsorbate groups. 
The latter demonstrates that there is also a decrease in the strain level, i.e. photoexcitation is reducing the strain on the graphene. This is a remarkable effect, which has not been reported before. From Fig.~\ref{fig2} it can be seen that photoexcitation also leads to broadening of distribution which for 1~$\textrm{mJ}/\textrm{cm}^2$ and 2~$\textrm{mJ}/\textrm{cm}^2$ excitation is mostly caused by variation of doping and for 3~$\textrm{mJ}/\textrm{cm}^2$ it is predominantly strain-induced.

\begin{figure}
\includegraphics{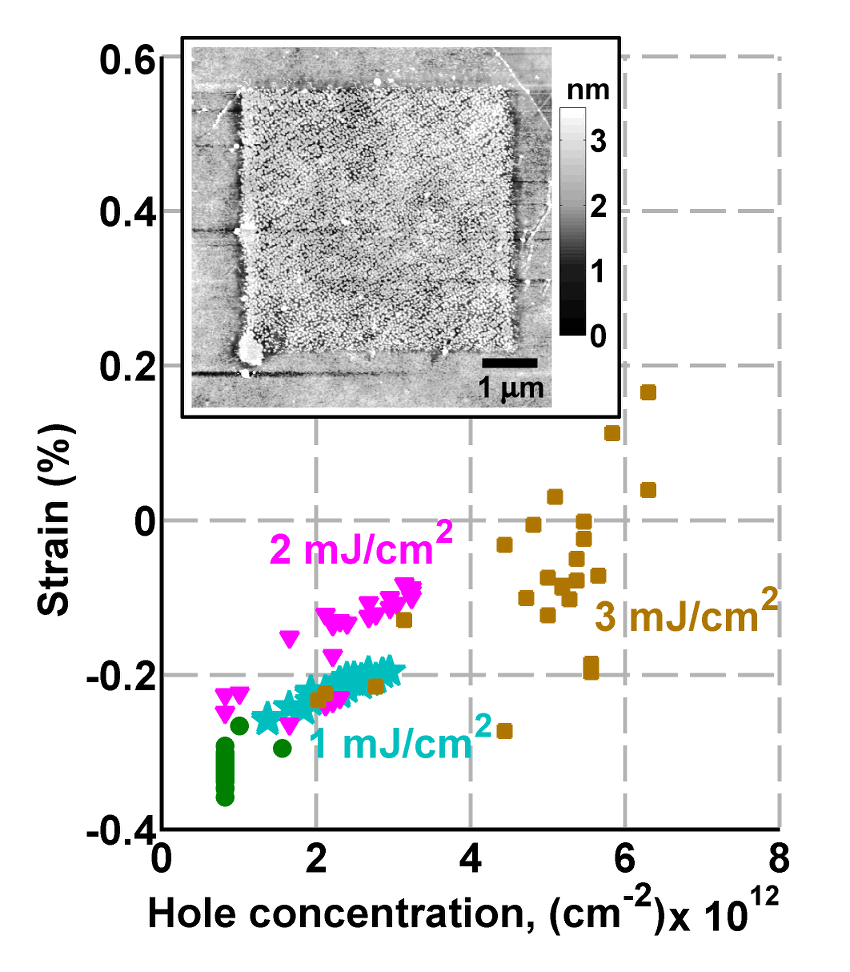}
\caption{\label{fig3}(Color online) Strain and carrier concentration induced by photoexcitation with 1~$\textrm{mJ}/\textrm{cm}^2$ (cyan stars), 2~$\textrm{mJ}/\textrm{cm}^2$ (magenta triangles) and 3~$\textrm{mJ}/\textrm{cm}^2$ (brown squares) laser pulses extracted from correlated peak positions. Inset: AFM image of a sample after photomodification.}
\end{figure}

The non-orthogonal coordinate system, used in Fig.~\ref{fig2}, complicates data interpretation. To clarify this, we can perform a vector decomposition to separate strain and doping contribution to the changes of the G peak position. We then use data from Refs.~\onlinecite{31Das2008}~and~\onlinecite{33Ni2008} to obtain carrier concentration and strain levels which correspond to the observed G peak shifts. Note that for hole concentrations $\textrm{n} < 1.4 \times 10^{12} \textrm{cm}^{-2}$ the linear approximation for doping-induced changes of $(\omega_G ; \omega_{2D})$ is no longer valid, making vector decomposition and data fitting ambiguous.  
From Fig.~\ref{fig3} it can be seen that photomodification leads to both increasing level of p-doping and reduction of compressive strain; higher laser fluences cause larger changes in doping and strain.

We have observed slightly differing magnitudes and shifts in a number of graphene flakes, most likely depending on the starting strain of the flake, though this has proven difficult to correlate. While the magnitudes of changes per unit excitation fluence vary from sample to sample, the sign of the changes is predominantly the same, resulting in an increase in hole doping and a reduction in compressive strain. This behaviour has been observed in five different samples. 
 
To conclude, we have investigated the modification of single layer graphene due to intense, picosecond near-infrared laser pulses. We find that photoexcitation leads to both a local increase of p-doping and reduction of compressive strain. With the short, intense laser pulses used in our experiments a number of mechanisms are feasible, including multi-photon excitation and non-equilibrium heating of the sample. The evidence from our experiments points towards enhanced atmospheric oxygen binding due to surface distortion, most likely caused by the rapid heating of the graphene. The inset of Fig.~\ref{fig3} shows an AFM image of one of the samples after modification, indicating the modified region has an increased surface roughness compared with surrounding non-modified areas. These changes are likely to be caused by slippage and buckling of the flake due to the mismatch in thermal expansion coefficients of graphene and underlying substrate. The doping level, meanwhile, can be explained by the enhanced bonding of atmospheric oxygen due to the distortion of graphene surface\cite{36Ryu2010}. These effects are similar to the compressive strain and p-doping introduced in graphene upon annealing\cite{35Lee2012, 36Ryu2010}, though the different result, i.e. the reduction of compressive strain, may be explained by the very local nature of heating for our experiments.  

Nevertheless, the local nature of the effects reported here could be utilized to create complex patterns that define device functionality, offering an advantage in spatial resolution and speed. However, to be able to change local properties of graphene in a controllable manner, more in-depth investigations to uncover the precise mechanisms at work are required.

\end{document}